\documentclass[twocolumn]{aastex6}

\usepackage{natbib,epsfig,graphicx,amssymb,epstopdf,url,bm,aasdefs}

\newcommand{\swtool}[1]{\textit{#1}}
\newcommand{\swpkg}[1]{\textsc{#1}}

\newcommand{\mcc}[2]{\multicolumn{#1}{c}{#2}}
\newcommand{\mccc}[1]{\multicolumn{1}{c}{#1}}

\newcommand{\mcr}[2]{\multicolumn{#1}{r}{#2}}
\newcommand{\mphn}{\phn\phn}

\newcommand{\xray}{\mbox{X-ray}}
\newcommand{\gray}{\mbox{gamma-ray}}

\newcommand{\uJy}{\mbox{$\mu$Jy}}
\newcommand{\ergcmsq}{\mbox{erg cm$^{-2}$}}
\newcommand{\ergcms}{\mbox{erg cm$^{-2}$ s$^{-1}$}}
\newcommand{\ergsec}{\mbox{erg s$^{-1}$}}
\newcommand{\kmsec}{\mbox{km s$^{-1}$}}
\newcommand{\pccmqb}{\mbox{pc cm$^{-3}$}}
\newcommand{\gpcyr}{\mbox{Gpc$^{-3}$ yr$^{-1}$}}
\newcommand{\rah}{\mbox{$^{\rm h}$}}
\newcommand{\ram}{\mbox{$^{\rm m}$}}

\newcommand{\rasxgd}[4]{\mbox{#1\rah\,#2\ram\,#3\fs #4}}
\newcommand{\dcsxg}[3]{\mbox{#1\arcdeg\,#2\arcmin\,#3\arcsec}}
\newcommand{\dcsxgd}[4]{\mbox{#1\arcdeg\,#2\arcmin\,#3\farcs #4}}

\newcommand{\xshooter}{\mbox{X-shooter}}
\newcommand{\usnob}{\mbox{USNO-B2.0}}
\newcommand{\frb}{\mbox{FRB\,131104}}
\newcommand{\swiftgrt}{\mbox{Swift~J0644.5$-$5111}}
\newcommand{\swiftxa}{\mbox{Swift~J064339.9$-$512042}}
\newcommand{\swiftxb}{\mbox{Swift~J064409.6$-$511853}}

\newcommand{\frbrpt}{\mbox{FRB\,121102}}
\newcommand{\frbag}{\mbox{FRB\,150418}}
\newcommand{\frbrot}{\mbox{FRB\,110523}}
\newcommand{\sgrbig}{\mbox{SGR\,1806$-$20}}

\begin{document}


\title{Discovery of a transient gamma-ray counterpart to FRB\,131104}

\author{J.~J. DeLaunay\altaffilmark{1,3,5},
        D.~B. Fox\altaffilmark{2,3,4},
        K. Murase\altaffilmark{1,2,3,4},
        P. M\'{e}sz\'{a}ros\altaffilmark{1,2,3,4},
        A. Keivani\altaffilmark{1,3},
        C. Messick\altaffilmark{1,3}, \\
        M.~A. Mostaf\'{a}\altaffilmark{1,3},
        F. Oikonomou\altaffilmark{1,3},
        G. Te\v{s}i\'{c}\altaffilmark{1,3}, and
        C.~F. Turley\altaffilmark{1,3}}

\affil{$^1$Department of Physics, Pennsylvania State University,
           University Park, PA 16802, USA \\
       $^2$Department of Astronomy \& Astrophysics, Pennsylvania
           State University, University Park, PA 16802, USA \\
       $^3$Center for Particle \& Gravitational Astrophysics,
           Institute for Gravitation and the Cosmos, Pennsylvania
           State University, University Park, PA 16802, USA \\
       $^4$Center for Theoretical \& Observational Cosmology,
           Institute for Gravitation and the Cosmos, Pennsylvania
           State University, University Park, PA 16802, USA}

\altaffiltext{5}{\href{mailto:jjd330@psu.edu}{\tt jjd330@psu.edu}}



\begin{abstract}

  We report our discovery in Swift satellite data of a transient
  \gray\ counterpart (3.2$\sigma$ confidence) to the fast radio burst
  \frb, the first such counterpart to any FRB. The transient has
  duration $T_{90} \simgt 100$\,s and fluence $S_\gamma\approx 4\times
  10^{-6}$\,\ergcmsq, increasing the energy budget for this event by
  more than a billion times; at the nominal $z\approx 0.55$ redshift 
  implied by its dispersion measure, the burst's \gray\ energy output
  is $E_\gamma \approx 5\times 10^{51}$\,erg.
  The observed radio to \gray\ fluence ratio for \frb\ is consistent
  with a lower limit we derive from Swift observations of another FRB,
  which is not detected in gamma-rays, and with an upper limit 
  previously derived for the brightest \gray\ flare from \sgrbig,
  which was not detected in the radio. 
  \xray, ultraviolet, and optical observations beginning two days
  after the FRB do not reveal any associated afterglow, supernova, or
  transient; Swift observations exclude association with the brightest
  65\% of Swift gamma-ray burst \xray\ afterglows, while leaving the
  possibility of an associated supernova at much more than 10\% the
  FRB's nominal distance, $D\simgt 320$\,Mpc, largely unconstrained. 
  Transient high-luminosity \gray\ emission arises most
  naturally in a relativistic outflow or shock breakout, as for
  example from magnetar flares, gamma-ray bursts, relativistic
  supernovae, and some types of galactic nuclear activity. Our
  discovery thus bolsters the case for an extragalactic origin for some
  FRBs and suggests that future rapid-response observations might
  identify long-lived counterparts, resolving the nature of these
  mysterious phenomena and realizing their promise as probes of
  cosmology and fundamental physics.

\end{abstract}



\section{Introduction}

Fast radio bursts (FRBs) are millisecond-long bursts of coherent
GHz-frequency emission \citep{lbm+07,tsb+13}, now regularly discovered
by radio pulsar surveys and survey facilities. Interest in this
population has been stimulated by their large dispersion measures,
${\rm DM} \simgt 300$\,pc cm$^{-3}$, which suggest an origin at
cosmological distances $D\simgt 1$\,Gpc (potentially in combination
with substantial plasma densities local to the source), and by their
high all-sky rate, estimated at $\mathcal{R} \approx 2100$\,day$^{-1}$
for fluences $S_{\rm GHz} > 2$\,Jy\,ms \citep{cpk+16}. Using
dispersion measures to infer distances in a standard cosmology (e.g.,
\citealt{ckw16}) gives a $z\approx 0.85$ horizon for FRB detection
with current facilities, yielding a lower bound on their volumetric
rate of 6700\,\gpcyr\ or 7\% the rate of core collapse supernovae
\citep{tcd+14}.

FRBs are thus a dramatic feature of the radio sky and an important
component of the transient activity of the local extragalactic or
cosmological universe. Yet despite intensifying efforts at real-time
discovery and follow-up \citep{pbb+15,rsj15,kjb+16}, along with
identification \citep{ssh+16a} and detailed studies \citep{ssh+16b} of
a single repeating source (\frbrpt), no non-radio counterpart or
high-confidence host galaxy for any FRB has been found, leaving their
distances, energy scales, and physical nature(s) unresolved.

In the absence of such counterparts, clues to the nature of the FRBs
have accumulated primarily via radio observations. Although only
\frbrpt\ is currently known to repeat, most of the fainter bursts from
this source would not have been detected at facilities other than
Arecibo. Further FRB repeaters may wait to be discovered
(e.g.\ FRBs\,110220 and 140514; \citealt{maoz+15}), though limits from
less sensitive facilities \citep{pjk+15} suggest they are likely a
minority.

The 44\% linear polarization of \frbrot\ enabled simultaneous
measurement of its dispersion and rotation measures, demonstrating the
presence of excess magnetized plasma along the line of sight, likely
located near the source in its external host galaxy
\citep{mls+15}. This has provided substantial support for cosmological
scenarios (the DM-based redshift estimate for \frbrot\ is $z\approx
0.5$), especially models with relatively young progenitors that would
be associated with nuclear or star-forming regions or a surrounding
supernova remnant \citep{mls+15,mkm16}.


One of five FRBs reported by \citet{cpk+16} exhibited a double-peaked
profile, with two peaks separated by $\Delta t\approx 5$\,ms
(FRB\,121002). This may disfavor catastrophic scenarios, e.g.\ binary
neutron star (BNS) mergers \citep{cpk+16}.


Rapid-response observations of \frbag\ \citep{kjb+16} across multiple
bandpasses identified a variable radio source, superposed on a
$z=0.49$ host galaxy, that was proposed as the fading afterglow of a
short-hard (BNS merger) gamma-ray burst. However, subsequent
observations revealed that the radio variable was in fact the galaxy's
active nucleus (AGN) rather than an afterglow \citep{wb16}, leaving
this FRB also without a high-confidence non-radio counterpart or host
galaxy. 


An FRB model invoking maser-like flaring of Galactic flare stars has
also been put forward \citep{lsm14}. Challenges for this model include
precisely reproducing the observed dispersion relation while avoiding
free-free absorption in a high-density setting \citep{kon+14,mkm16},
an absence of increased FRB rates toward the Galactic plane
\citep{cpk+16}, limits on source repetition \citep{pjk+15}, and the
absence of known or apparent variable stars in association with most
known FRBs \citep{maoz+15}.


In this paper we present a search for untriggered (subthreshold)
\gray\ FRB counterparts. This is not the first search for
\gray\ counterparts to FRBs. Following identification of \frbrpt\ as a
repeating source, \citet{ssh+16b} reviewed Swift BAT above-threshold,
Fermi Gamma-ray Burst Monitor (GBM; \citealt{fermi_gbm}) subthreshold,
and Fermi Large Area Telescope (LAT; \citealt{fermi_lat}) subthreshold
datasets without identifying any significant transient \gray\ activity
from that source.

\citet{tkp16} considered a set of during-, pre-, and post-FRB
\gray\ observations from the Swift BAT, Fermi GBM, and Konus-Wind
\citep{konuswind} instruments, without identifying any likely
\gray\ counterparts, and derived the first limits on any FRB-like
counterpart to the 2004 December 27 giant \gray\ flare from \sgrbig.

Taking an alternate approach, \citet{bmg+12} implemented a program of
rapid-response radio observations of gamma-ray bursts (GRBs) using the
Parkes Observatory 12\,m dish and reported two candidate associated
radio bursts (for GRBs 100704A and 101011A) from nine observed
GRBs. However, these candidates may be artifacts of radio frequency
interference; null results from a more sensitive subsequent search by
\citet{pwt+14}, and other GRB rapid-response experiments discussed
therein, lend weight to this interpretation.

Going forward, the promise of VLA or other interferometric detections
of FRBs is substantial \citep{lbb+15}, as these would yield
sub-arcsecond positions from the burst data alone. Such positions
could yield high-confidence host galaxy identifications without the
need to identify non-radio transient counterparts.


Apart from hopes that FRB counterparts or host galaxies will finally
reveal the physical nature(s) of these sources, either counterparts or
precise distances will be required if FRBs are to fulfill their
substantial promise as probes of cosmology
\citep{ioka03,inoue04,arg16} and fundamental physics
\citep{wgw+15,wzg+16}.


Our manuscript proceeds as follows: We detail the search which yielded
discovery of the gamma-ray counterpart to \frb\ in Sec.~\ref{sec:obs},
along with our analysis of relevant archival and follow-up
observations. In Sec.~\ref{sec:discuss} we explore possible
interpretations of our findings, considering FRB models and various
high-energy transient source populations. We conclude in
Sec.~\ref{sec:conclude}.


\section{Observations and Analysis}
\label{sec:obs}


\subsection{Swift Subthreshold Search}
\label{sub:search}


We carried out a search for untriggered (subthreshold) transient
\gray\ counterparts to all FRBs from the \textsc{frbcat}
catalog\footnote{\textsc{frbcat}:
  \url{http://www.astronomy.swin.edu.au/pulsar/frbcat/}}
\citep{frbcat}, including all reported bursts from the repeating
\frbrpt. We examined \gray\ data from the Swift Burst Alert Telescope
(BAT; \citealt{bat2005}), in near-continuous operation since November
2004, and from the International Gamma-Ray Astrophysics Laboratory's
IBIS imager \citep{ibis2003}, in near-continuous operation since
October 2002.

During this time, two of 13 non-repeating FRBs and two of 17 bursts
from \frbrpt\ occurred within the BAT field of view; no FRBs occurred
within the IBIS field of view. For each FRB with simultaneous BAT
coverage, we retrieved the relevant data from the High-Energy
Astrophysics Science Archive Research Center
(HEASARC\footnote{HEASARC:
  \url{http://heasarc.gsfc.nasa.gov/}}) and searched for
sources within 15$\arcmin$ of the FRB coordinates.  This radius
accounts for uncertainty in the positions of both the radio source
(typically localized to a single beam with ${\rm FWHM} \approx
15\arcmin$) and the subthreshold BAT source candidates (having
90\%-containment radii $r_{90}\approx 7\arcmin$) that are observed in
these data.

We used the \swpkg{heasoft} (v.\ 6.18) software tools and calibration
for our high-energy data analyses\footnote{\textsc{heasoft}:
  \url{http://heasarc.nasa.gov/lheasoft/}}. Swift BAT survey data
include detector plane histograms (DPHs) of the full-bandpass
(15--195\,keV) 300\,s exposures and scaled detector plane images
(DPIs) of the soft-band (15--50\,keV) 64\,s exposures. We reduced
these data using standard procedures, adopting the maximum allowed
oversampling parameter of 10, and searched for candidate sources using
the \swtool{batcelldetect} sliding-cell algorithm. This routine uses
local estimates of the background and noise level to identify
candidate sources, and then performs a point-spread function (PSF) fit
to derive an accurate source position and BAT counts estimate. We
estimated uncertainties in source positions ($r_{90}$) from source
significances using the calibration of \citet{batsurvey70}\ (their
Eq.~7).


As we are interested in testing the hypothesis of a fixed
$S_\gamma:S_{\rm GHz}$ fluence ratio for FRBs -- and as we are
interested in non-repeating sources (as candidate catastrophic events)
more than in the known repeating source \frbrpt\ -- we prioritized the
search as follows: non-repeating FRBs ordered by decreasing radio
fluence, followed by bursts of \frbrpt\ ordered by decreasing radio
fluence. The results of our search are presented in
Table~\ref{tab:batprompt}.


\begin{table*}
  \footnotesize
  
  \caption{\label{tab:batprompt}%
    Swift BAT observations of FRBs}

  \begin{center}    

    \begin{tabular}{c r r c c c c c} 

    \hline \hline

 \mcc{1}{FRB} 
     & \mcc{1}{R.A.} 
     & \mcc{1}{Dec.}
     & $S_{\rm GHz}$ 
     & UTC 
     & BAT $\Delta t$
     & $S_\gamma$ 
     & $\log_{10} \eta$ \\

  ~ & ~ & ~ 
     & (Jy\,ms) & ~
     & (s) 
     & ($10^{-6}$\,\ergcmsq)
     & ~ \\
 \hline

 131104 & 101.062 & $-$51.278 & 2.33 & 2013-11-04 18:03:59 & $-$7, +293 & 4.0\,$\pm$\,1.8 & 5.8\,$\pm$\,0.2 \\

           ''  & \mcc{1}{''}  & \mcc{1}{''} & '' & '' & +293, +593 & $<$2.8 & n/a \\

 110626 & 315.929 & $-$44.739 & 0.56 & 2011-06-26 21:33:15 & $-$84, +216 & $<$1.1 & $>$5.7 \\

 121102 (2) & ~82.992 & 33.134 & 0.11  & 2015-05-17 17:42:09 & $-$46, +254 & $<$2.4 & $>$4.6 \\

 121102 (3) & ~82.992 & 33.134 & 0.10 & 2015-05-17 17:51:41 & $-$18, +248 & $<$2.0 & $>$4.7 \\ \hline

\multicolumn{8}{p{5.9in}}{\scriptsize%
  {\bfseries Note---}FRB radio properties from \textsc{frbcat}
  \citep{frbcat}. BAT pointing $\Delta t$ (start, stop) intervals are
  given with respect to the quoted topocentric FRB time;
  \gray\ fluences $S_\gamma$ (15--150 keV) are calculated using the
  best-fit photon index $\Gamma=1.16$ power-law spectrum for \frb\ and
  assuming a photon index $\Gamma=2$ power-law spectrum for the other
  events; radio to \gray\ fluence ratios $\eta \equiv S_{\rm
    GHz}/S_{\gamma}$ are in units of Jy\,ms\,erg$^{-1}$\,cm$^2$. BAT
  survey image ObsIDs are: 00571830037 (\frb), 00040453002
  (FRB\,110626), and 00036376034 (for both bursts of \frbrpt). Further
  details on FRB radio properties are available from
  \textsc{frbcat}.}

 \end{tabular}
 \end{center}
\end{table*}



\subsection{Counterpart Discovery}
\label{sub:discovery}

\begin{figure}
\centerline{\includegraphics[width=89.0mm]{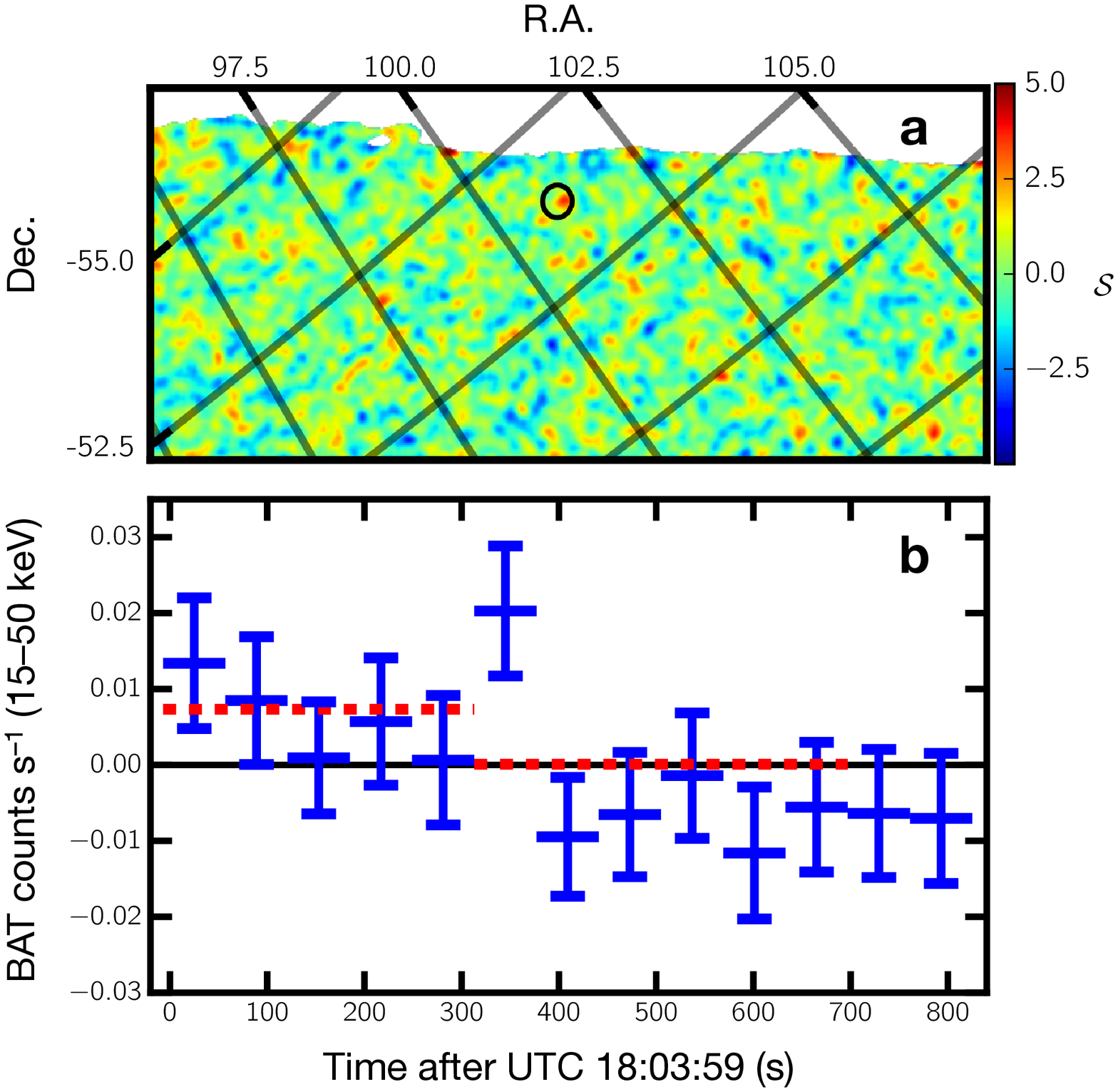}}
\vspace*{\baselineskip}
\caption{\footnotesize Swift BAT discovery image and light
    curve for the transient \gray\ counterpart to \frb, \swiftgrt.
  (a) \swiftgrt\ discovery image (15--150\,keV; UTC 18:03:52 start;
  300\,s exposure), showing a small portion of the BAT field of
  view in tangent plane projection. The search region for \frb\ (black
  circle) is shown; regions with $<$1\% coding are masked. The
  point-like excess associated with the \gray\ transient peaks at
  signal-to-noise $\mathcal{S} = 4.2\sigma$.
  (b) Soft-band (15--50\,keV) light curve for \swiftgrt. Time is
  measured from the FRB detection, UTC 18:03:59. Both 64\,s (blue) and
  320\,s (red dashed) flux measurements are shown; error bars are
  $\pm$1$\sigma$.
  \label{fig:loclc}}
\end{figure}

We identified an untriggered \gray\ transient candidate with
signal-to-noise $\mathcal{S}=4.2\sigma$ in the first search area, that
associated with \frb\ \citep{rsj15}. The transient position is
R.A. \rasxgd{06}{44}{33}{12}, Dec.\ \dcsxgd{$-$51}{11}{31}{2} (J2000),
with $r_{90}=6\farcm 8$ (Fig.~\ref{fig:loclc}). It is located near the
edge of the BAT field of view, with only 2.9\% of BAT detectors
illuminated through the coded mask (2.9\% coding), which explains its
low significance in spite of a relatively bright inferred fluence. Its
sky position is well within the search area, 6\farcm 3 from the radio
receiver pointing, with 50\% of its BAT localization probability
within the receiver FWHM (Fig.~\ref{fig:sky}). No candidate
counterparts are identified for the remaining FRBs with BAT coverage,
with results as reported in Table~\ref{tab:batprompt}.

Since a \gray\ transient is identified for the highest radio fluence
non-repeating FRB in our sample, and since the $S_\gamma:S_{\rm GHz}$
constraints for the other FRBs are consistent with the ratio inferred
for \frb, this is consistent with our hypothesis and first test, and
we adopt a trials factor of one for assessing the significance of the
counterpart.


We determine this significance by examining 1429 archived BAT survey
pointings with exposure times 200\,s to 400\,s that were taken over
the one-year period June~2015 to May~2016. On average each of these
survey images has 46.3 transient candidates with $\mathcal{S} \geq
4.2\sigma$ at $>$1\% coding; although some may be cosmic sources, for
present purposes we treat them all as noise fluctuations. The density
of candidates per unit solid angle varies across the field of view, so
we focus on a rectangular region of the BAT image plane, centered on
the transient position in tangent plane coordinates. Within this
region, which has area 16\,deg$^2$ (0.36\% of the field of view), we
find an average of $0.106\pm 0.009$ transient candidates per survey
pointing. This average density of candidates does not vary
systematically with pointing coordinates, exposure time, or date, over
the set of BAT pointings.  The resulting $p$-value, corresponding to
the number of expected candidates in a single 15\arcmin\ search
radius, is $p=0.13\%$, corresponding to Gaussian confidence level
$\mathcal{C} = 3.2\sigma$.


We also estimate the significance of the association by an alternate
Bayesian approach that does not require us to define a search area in
advance. Instead, we make a point comparison of the probability
density for an FRB-associated transient (at the maximum-likelihood
position of the \gray\ transient) to the probability density for a
noise fluctuation. The former will be distributed according to the FRB
positional uncertainty (derived below), while the latter will be
uniform over the larger rectangular test region. Because the transient
is located 1.2$\sigma$ from the FRB coordinates (distance
$d=6.25\arcmin$, $\sigma=5.2\arcmin$), the FRB counterpart probability
density is $2.85\times 10^{-3}$\,arcmin$^{-2}$; while the background
source density is $1.84\times 10^{-6}$\,arcmin$^{-2}$. Hence an
association is preferred by an odds ratio of 1552:1, corresponding to
Gaussian confidence level $\mathcal{C}=3.4\sigma$.

As both metrics exceed the 3$\sigma$ threshold common for counterpart
identification in astrophysics, we consider the transient confirmed
and designate it \swiftgrt, the first non-radio counterpart to any
FRB. Its properties are summarized in Tables~\ref{tab:batprompt} and
\ref{tab:props}.


\begin{table}
  \footnotesize

  \caption{\label{tab:props}%
    Properties of \frb}

  \begin{center}
    
    \begin{tabular}{l l r}
      \hline \hline
Joint & R.A. & \rasxgd{06}{44}{27}{06} \\
~      & Dec. & \dcsxgd{$-$51}{12}{54}{0} \\
~       & $r_{90}$ & 5\farcm 78 \\ \hline

Radio & UTC & 18:03:59 \\
~        & $S_{\rm GHz}$ & 2.33 Jy ms \\
~        & DM & 779\,$\pm$\,1\,\pccmqb \\
~        & $z_{\rm max}$ & 0.55 \\ \hline

$\gamma$-ray  & $T_{90}$ & 377\,$\pm$\,24\,s (1$\sigma$) \\  
   ~          & \mcr{2}{$>$100\,s (90\%-c.l.)} \\

\mcr{1}{PL} & $\Gamma$ & 1.16$^{+0.68}_{-0.78}$ \\
~         & $S_{\gamma,-6}$ & 4.0\,$\pm$\,1.8 \\ 

\mcr{1}{TB} &  $kT$ & 200$^{+\infty}_{-125}$\,keV \\
~        & $S_{\gamma,-6}$ & 3.4\,$\pm$\,1.5 \\ \hline

\multicolumn{3}{p{2.8in}}{\scriptsize {\bfseries Note---}Radio
  properties including topocentric burst time (UTC), radio fluence
  ($S_{\rm GHz}$), and dispersion measure (DM) are from \citet{rsj15},
  while the maximum redshift for a consensus cosmology ($z_{\rm max}$)
  is from \citet{mkm16}. $\gamma$-ray and joint properties are from
  this work. Coordinates for the joint radio + \gray\ localization are
  J2000. Spectral parameters for power-law (PL) and thermal
  bremsstrahlung (TB) fits are quoted with 90\%-confidence intervals;
  \gray\ fluences $S_{\gamma,-6}$ (15--150 keV) are in units of
  $10^{-6}$\,\ergcmsq.}

  \end{tabular}
  \end{center}
\end{table}


\subsection{Counterpart Properties}
\label{sub:tprops}


We first seek to refine the location for \frb\ by combining radio and
\gray\ constraints. This requires a quantitative form for the radio
localization; since this has not previously been derived in a rigorous
fashion, we describe our approach here.

\begin{figure}
  \centerline{\includegraphics[width=74.0mm]{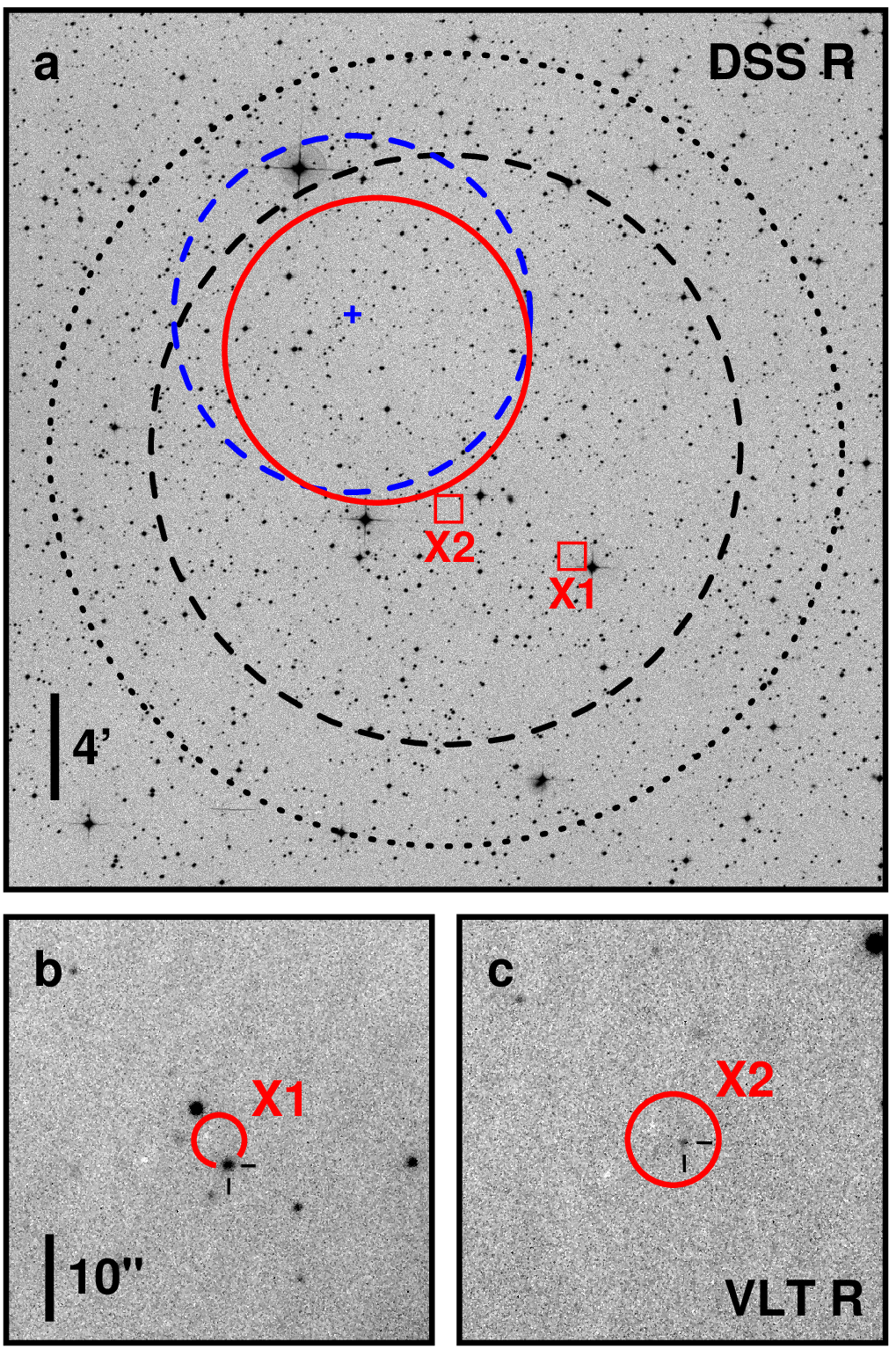}}
  \caption{\footnotesize Localizations for \frb\ and its
      transient \gray\ counterpart, \swiftgrt.
    (a) Archival $R$-band image of the search region ($R=15\arcmin$,
    black dotted circle) and radio localization ($r_{90} = 11\farcm
    2$, black dashed circle) for \frb; \gray\ localization
    ($r_{90}=6\farcm 8$, blue dashed circle, centered on blue +) for
    \swiftgrt; and resulting joint radio + \gray\ localization
    ($r_{90}=5\farcm 8$, red solid circle).  Positions of the two
    identified Swift \xray\ sources are indicated (red squares).
    (b) \xshooter\ $R$-band image of the \swiftxa\ region, showing its
    optical counterpart (ticks), a $z=0.383$ quasar with
    $R\approx 19.0$\,mag. 
    (c) \xshooter\ $R$-band image of the \swiftxb\ region, showing its
    optical counterpart (ticks), a $z=1.525$ quasar with
    $R\approx 20.8$\,mag.  
    \label{fig:sky}}
\end{figure}

We assume an azimuthally-symmetric Gaussian form for the response of
the Parkes multibeam receiver \#5, which has a quoted FWHM of
$15\arcmin$ \citep{rsj15}. In parallel, we assume an $N(S_{\rm
  GHz}>S_0) \propto S_0^{-3/2}$ form for the radio fluence
distribution of FRBs \citep{lbb+15}. Together, these assumptions imply
a two-dimensional Gaussian probability distribution for the location
of \frb, with $\sigma=5.2\arcmin$ and 90\%-containment radius
$r_{90}=11\farcm 2$, centered on the receiver pointing coordinates
R.A. \rasxgd{06}{44}{10}{40}, Dec. \dcsxg{$-$51}{16}{40} (J2000). This
puts 92\% of the BAT localization within the radio-derived
$r_{90}$. Weighting this localization with the $r_{90}=6\farcm 75$
localization of the \gray\ transient yields a joint localization
centered at R.A. \rasxgd{06}{44}{27}{06},
Dec.\ \dcsxgd{$-$51}{12}{54}{0} (J2000), with $r_{90}=5\farcm
78$. This localization is illustrated in Fig.~\ref{fig:sky} and
reported in Table~\ref{tab:props}.


Examination of archival images of the burst and transient localization
region (Fig.~\ref{fig:sky}) does not reveal any prominent Local Group
or low-redshift galaxies, nor bright active galaxies, although as
noted by \citet{rsj15}, the field is near the projected tidal limit of
the Carina dwarf spheroidal galaxy ($D\approx 100$\,kpc) and a
projected tidal stream of the Large Magellanic Cloud ($D\approx
50$\,kpc). The absence of known or candidate flare stars has been
noted by \citet{maoz+15}.


A NASA/IPAC Extragalactic Database (NED\footnote{NED:
  \url{https://ned.ipac.caltech.edu}}) query targeting the predefined
search area for \frb\ yields six cataloged galaxies\footnote{Galaxies
  within the \frb\ radio localization: GALEXASC~J064303.18$-$511832.0,
  2MASX~J06430652$-$5110339, 2MASX~J06434024$-$5113110,
  \mbox{ESO~206-G 022}, 2MASX~J06435104$-$5110507,
  2MASX~J06435472$-$5118337.}, the quasar QJ0643$-$5126 ($z=2.77$),
and the IRAS source IRAS F06441$-$5118. The observed density of
resolved galaxies and the presence of a known quasar and an IRAS
source are not remarkable for a field of this size at this Galactic
latitude. All of these cataloged sources lie well outside the joint
localization region except for 2MASX~J06435104$-$5110507, which at
6\arcmin\ distance from the center of the joint localization lies just
outside its $r_{90}$.


We generated a spectrum of \swiftgrt\ from the 300\,s detection image
and fitted spectral models within the \swpkg{xspec} environment. The
relatively low signal-to-noise admits a broad range of spectral
models, including simple power-law and thermal bremsstrahlung models,
which we prefer and present in Tables \ref{tab:batprompt} and
\ref{tab:props}. Using the best fit power-law we derive a fluence of
$S_\gamma = 4.0\pm 1.8 \times 10^{-6}$\,\ergcmsq\ (15--150\,keV).


This implies a radio to \gray\ fluence ratio of $\log\eta = 5.8\pm
0.2$ for \frb, where $\eta \equiv S_{\rm GHz}/S_\gamma$ is expressed
in units of Jy ms erg$^{-1}$ cm$²$ as defined by \citet{tkp16}. Those
authors estimate that the \sgrbig\ giant flare had $\log\eta < 5.9$
based on modeling of the sidelobe response of the Parkes multibeam
receiver (strong constraint), or alternatively, $\log\eta < 7.9$ in an
idealized diffraction-limited treatment (weak constraint). Our value
for \frb, consistent with the lower limits we derive for FRB\,110626
and for two fainter bursts from \frbrpt\ (Table~\ref{tab:batprompt}),
is consistent with these upper limits from
\sgrbig\footnote{\citet{tkp16} derive lower limits on $\log\eta$ for a
  number of FRBs, including \frb, that are inconsistent with their
  \sgrbig\ limit. However, they allow any observation of the FRB
  position within 10~minutes before or after the burst to limit its
  \gray\ fluence, and use a \gray\ duration of $T_{90}=0.1$\,s to
  derive limits from non-imaging instruments, the Fermi GBM and
  Konus-Wind. Their limits are thus not directly comparable with ours,
  since we analyzed only simultaneous observations by an imaging
  instrument, the Swift BAT.}.

For a nominal 10~GHz bandpass and the observed flat or inverted
spectrum \citep{rsj15}, the integrated radio fluence of \frb\ is
$S_{\rm radio} \sim 3\times 10^{-16}$\,\ergcmsq. The
\gray\ counterpart thus increases the energy requirements for this
event by a factor of roughly ten billion. (Given its inferred
off-center location within the receiver beam, the radio fluence of
\frb\ may be underestimated.)


We constructed a light curve of \swiftgrt\ using
\swtool{batcelldetect} to measure the counts from \swiftgrt\ in each
of thirteen 64\,s and two 320\,s soft-band (15--50\,keV) exposures
covering the sky position of the transient during and immediately
subsequent to the occurrence of \frb. The soft-band light curve,
presented in Fig.~\ref{fig:loclc} and Table~\ref{tab:batlc}, suggests
a transient duration (for 90\% of the burst fluence) of $T_{90}\approx
380$\,s, extending partway into the next 300\,s exposure, which
maintained the same pointing.


\begin{table}
\footnotesize

 \caption{\label{tab:batlc}%
    BAT light curve for \swiftgrt}

 \begin{center}
 \begin{tabular}{r r r r}
   \hline \hline

   $t_{\rm start}$ & 
   Exp. &
   Flux &
   Unc. \\ 

   \mcc{1}{(s)} & 
   \mcc{1}{(s)} & 
   \mcc{2}{(cts ks$^{-1}$)} \\ \hline

   $-$7  & 320 & 7.34 & 4.26 \\ 
   $-$7  & 64 & 13.40 & 8.61 \\
   57   & 64 & 8.48 & 8.41 \\
   121 & 64 & 0.93 & 7.38 \\
   185 & 64 & 5.72 & 8.38 \\
   249 & 64 & 0.62 & 8.53 \\ 
   313 & 320 & 0.09 & 3.80 \\ 
   313 & 64 & 20.28 & 8.56 \\
   377 & 64 & $-$9.47 & 7.82 \\
   441 & 64 & $-$6.54 & 8.19 \\
   505 & 64 & $-$1.41 & 8.26 \\
   569 & 64 & $-$11.59 & 8.69 \\ 
   633 & 64 & $-$5.57 & 8.54 \\
   697 & 64 & $-$6.39 & 8.43 \\
   761 & 64 & $-$7.04 & 8.57 \\  \hline

\multicolumn{4}{p{1.7in}}{\scriptsize {\bfseries Note---}BAT survey data
  provide two 320\,s integrations and thirteen 64\,s integrations over
  the 15--50\,keV bandpass for the pointings covering and immediately
  subsequent to \frb. Start times $t_{\rm start}$ are measured from
  the burst time, UTC 2013-11-04 18:03:59. Count rate uncertainties
  are 1$\sigma$.}

  \end{tabular}
  \end{center}
\end{table}


  
To quantify the likely burst duration, we adopt a simple ``step
function'' model, assuming a fixed active flux level, a start time
$\Delta t=0$ relative to the FRB, and an end time of $\Delta
t=T_{90}$. (We take this approach because the duration of a
step-function light curve more closely corresponds to $T_{90}$ than to
$T_{100}$ for realistic \gray\ transient light curves.) We make a
$\chi^2$ minimization of this function against the light curve at
64\,s resolution, finding a best-fit for $T_{90}=377$\,s ($\chi^2 =
9.9$ for 11 d.o.f.) which extends exactly to the end of the
highest-flux sixth time bin, and a 1$\sigma$ uncertainty range of
$T_{90} = 377\pm 24$\,s (Table~\ref{tab:props}).

We derive a model-independent lower bound on the burst duration by
noting that the transient signal-to-noise $\mathcal{S} \propto
S_\gamma/\sqrt{T_{90}}$ for relatively plateau-like light
curves. Since the transient has $\mathcal{S}=4.2\sigma$ over 300\,s,
for $T_{90} \le 125$\,s it would have $\mathcal{S} \ge 6.5\sigma$,
leading to a BAT trigger. This is because the BAT flight software
continuously evaluates count rates across the detector in search of
excesses (leading to a rate trigger, synthesis of a sky image, and a
search for a new point source), and synthesizes sky images at 64\,s
intervals and multiples thereof during each fixed pointing, in search
of new point sources (leading to an image trigger). Due to uncertainty
as to the exact shape of the light curve and its timing with respect
to BAT integration intervals, we conservatively quote $T_{90} >
100$\,s as our 90\%-confidence lower limit on the burst
duration. Since we cannot exclude the possibility of a very extended
duration at a very low flux level, we do not set a 90\%-confidence
upper bound. If the \gray\ emission did extend to $t > 293$\,s then
our quoted fluence will be an underestimate; including the additional
fluence seen in the sixth time bin increases the \gray\ fluence by
+66\%.

Given its duration ($T_{90} > 100$\,s), its spectrum, and the absence
of a BAT trigger, non-detections of the source by the Fermi GBM,
INTEGRAL SPI-ACS, Konus-Wind, and other Interplanetary Network
detectors are not further constraining of the counterpart's
\gray\ properties.


Given its observed dispersion measure, ${\rm DM} = 779 \pm
1$\,\pccmqb, the maximum distance for \frb\ is $D\approx 3.2$\,Gpc or
$z\approx 0.55$ \citep{mkm16}. This distance estimate will hold
approximately in the absence of significant plasma density near the
source, and will be reduced if there is any substantial quantity of
such local plasma. It is also subject to systematic uncertainties
regarding the ionized fraction of gas in the intergalactic medium. At
the maximum distance, the implied \gray\ energy output for \frb\ is
$E_\gamma \approx 5\times 10^{51}$\,erg.


\subsection{Swift Follow-Up}
\label{sub:swift}

The large inferred \gray\ energy release for \frb\ suggests the
possibility of subsequent high-energy afterglow, supernova, or
galactic nuclear emissions. We therefore reviewed a set of Swift
\xray\ and UV/optical observations of the burst position that were
taken two days after the original radio discovery, likely in a search
for such counterparts (Table~\ref{tab:followup}).


\begin{table*}
  \footnotesize
  
  \caption{\label{tab:followup}%
    Follow-Up Observations of \frb}

  \begin{center}    

    \begin{tabular}{lcrrrclp{1.8in}}

    \hline \hline

       \mccc{Instrument}
     & \mccc{Band} 
     & \mccc{$t_{\rm start}$} 
     & \mccc{$t_{\rm stop}$}
     & \mccc{Exp.}
     & \mccc{Cover}
     & \mccc{Limit}
     & \mccc{Comments} \\

     ~ & ~ & \mccc{(d)} & \mccc{(d)} & \mccc{(s)} & ~ & ~ & ~ \\ \hline

 Swift XRT & 0.3--10\,keV & 2.00 & 2.10 & 4912 & 97\% & $4\times 10^{-14}$
       & Excludes brightest 65\% of Swift GRB afterglows \\

 Swift UVOT & $U$  & 2.00 & 2.07 &  157 & 75\% & \phn 32\,\uJy 
       & $U>19.4$\,mag \\

 \mccc{''} & $B$  & 2.00 & 2.07 &  157 & \mccc{''} &\phn  59
       & $B>19.6$\,mag \\
    \mccc{''} & $V$  & 2.01 & 2.08 &  157 & \mccc{''} &     132 
       & $V>18.6$\,mag \\
    \mccc{''} & UVM2 & 2.01 & 2.10 & 3432 & \mccc{''} &\mphn 4.3 & ~ \\
    \mccc{''} & UVW1 & 2.00 & 2.14 &  322 & \mccc{''} &\phn 17   & ~ \\
    \mccc{''} & UVW2 & 2.01 & 2.08 &  629 & \mccc{''} &\mphn 8.1 & ~ \\

 VLT \xshooter & $R$ & 2.48 & 2.55 & 720 & 1\% & \mphn 4.5\,\uJy
       & Targeting X1 and X2; not used for transient search \\ \hline

\multicolumn{8}{p{5.84in}}{\scriptsize%
  {\bfseries Note---}Coverage fractions (``Cover'') are quoted for the
  \frb\ radio localization 90\%-confidence region ($r_{90}$), requiring
  achieved depth of $>$90\% of the exposure time for Swift UVOT and
  $>$50\% of the exposure time for Swift XRT; coverage fractions for
  the joint localization 90\%-confidence region are 96\% (Swift XRT),
  92\% (Swift UVOT), and 0\% (VLT \xshooter), respectively. The Swift
  XRT \xray\ limit is in units of \ergcms. UVOT images were searched
  for new sources by comparison to archival images; absence of deep
  pre-imaging forestalls a similar search with \xshooter\ data,
  which in any case do not provide coverage of the joint localization
  region.}

 \end{tabular}
 \end{center}
\end{table*}



We processed the Swift \xray\ Telescope (XRT; \citealt{xrt2005})
follow-up data (Swift ObsID 00033033001) with \swtool{xrtpipeline},
which provided a clean event list, image, and exposure map for the
observations. The XRT data cover 97\% of the \frb\ radio localization
(and 96\% of the joint localization) to better than half the nominal
4900\,s depth, and were taken in two extended exposures between
$t+2.0$\,days and $t+2.1$\,days post-burst. We identified sources
using the \swtool{wavdetect} routine from the \textsc{ciao} software
package\footnote{\textsc{ciao}: \url{http://cxc.harvard.edu/ciao/}},
running against a range of scales, setting the single-trial source
threshold at $p=10^{-6}$, and fixing the PSF FWHM at 12\arcsec.

This analysis yields two high-confidence \xray\ sources
(Fig.~\ref{fig:sky}) which we designate \swiftxa\ or X1 (located well
outside the joint localization) and \swiftxb\ or X2 (located
49\arcsec\ beyond the joint localization $r_{90}$, on the edge of the
95\%-containment region). For each of these sources, we produced
refined positions and $r_{90}$ estimates, optimally-extracted source
and background counts, and \xray\ spectral fits and flux estimates
using the routines of \citet{xrtauto2009} via the UK Swift Science
Data Centre\footnote{Build Swift-XRT Products:
  \url{http://www.swift.ac.uk/user_objects/}}.

These established that \swiftxa\ or X1 (significance
$\mathcal{S}=5.2\sigma$; 17 counts compared to 1.3 expected from
background), is located at R.A. \rasxgd{06}{43}{39}{96},
Dec.\ \dcsxgd{$-$51}{20}{42}{9} (J2000), with $r_{90}=3\farcs 1$, and
has an estimated \xray\ flux of $1.4^{+1.2}_{-0.7}\times
10^{-13}$\,\ergcms\ (0.3--10\,keV, 90\%-confidence bounds); while
\swiftxb\ or X2 ($\mathcal{S}=3.2\sigma$; 9 counts compared to 0.8
expected from background), is located at R.A. \rasxgd{06}{44}{09}{65},
Dec.\ \dcsxgd{$-$51}{18}{53}{6} (J2000), with $r_{90}=5\farcs 6$, and
has an estimated \xray\ flux of $8.0^{+13.5}_{-4.9}\times
10^{-14}$\,\ergcms. In separate analyses, we confirmed that the
arrival times and radial distributions of counts for both sources are
consistent with a non-variable point-source nature for each.


Given that X1 is located well outside the joint localization, and that
the $z=1.525$ redshift for X2 (derived below) is beyond the horizon
for \frb, we do not consider either source a viable counterpart. We
conclude instead that we have established an upper limit of $4\times
10^{-14}$\,\ergcms\ (0.3--10\,keV) on the flux of any
\xray\ counterpart at $t+2$\,days post-burst.

To determine a constraint on GRB afterglow-like counterparts, we
considered a library of 192 Swift-detected GRB \xray\ afterglows
analyzed and modeled by \citet{rlb+09}. Using the power-law decays of
these afterglows we interpolated or extrapolated the observed
\xray\ behavior to $t+2$\,days post-burst, finding that 65\% (125 of
192) of these afterglows have inferred fluxes above our limit. Hence
we conclude that the XRT observations exclude association of
\frb\ with the brightest 65\% of Swift-type \xray\ afterglows.


We processed Swift UV/Optical Telescope (UVOT; \citealt{uvot2005})
data with \swtool{uvotskycorr} and \swtool{uvotimsum}, which yields
aspect-corrected coadded images covering 75\% of the radio
localization (and 92\% of the joint localization) to better than 90\%
of the nominal exposures in each of six UVOT filters: $U$ (157\,s
exposure), $B$ (157\,s), $V$ (157\,s), UVM2 (3432\,s), UVW1 (322\,s),
and UVW2 (629\,s). We coadd these six images in a search for new or
anomalously bright sources by comparison to archival images; none are
found within the UVOT area. Using \swtool{uvotsource}, we derive
per-filter flux density upper limits for source-free regions of the
image as listed in Table~\ref{tab:followup}; these serve as upper
limits for any new point source in the UVOT region.

Review of the Swift GRB Table\footnote{Swift GRB Table:
  \url{http://swift.gsfc.nasa.gov/archive/grb_table/}} shows that
there are no examples of Swift burst afterglows detected by UVOT and
undetected by XRT; hence we prefer the XRT constraint (excluding the
brightest 65\% of Swift afterglows) for afterglow-like counterparts.


With respect to supernova (SN) limits, pervasive line-blanketing by
metals suppresses SN optical/UV flux at $\lambda < 4000\AA$, so that
the $V>18.6$\,mag and $B>19.6$\,mag limits (Table~\ref{tab:followup})
are the most useful for SN constraints. Observations at $t+2$\,days
post-explosion are sub-optimal for catching associated SNe, as maximum
light is achieved after two to three weeks, depending on SN type and
explosive energy. Tabulations of SN lightcurves suggest that optical
magnitudes at $t+2$ days are $\Delta M_V \approx 2$\,mag fainter than
peak for type Ibc supernovae \citep{dsg+11}, and likely even further
suppressed for type~Ia events \citep{fsg+15}, although explosion times
for Ia events are not well constrained. As a result, the UVOT limits
(ignoring unobserved portions of the joint localization, the 10\%
probability of the source lying outside the joint localization, and
any possible extinction in the host galaxy) are not constraining for
the nominal distance $D=3.2$\,Gpc, as they requires any associated SN
to have peak absolute magnitude $M_V > -23.9$\,mag ($M_B >
-22.9$\,mag).
UVOT limits become constraining at closer distances; at
$D=320$\,Mpc the limits of $M_V > -18.9$\,mag ($M_B > -17.9$\,mag) at
peak exclude the brightest $\approx$10\% of type Ibc supernovae
\citep{dsg+11} and the brightest $\approx$60\% of type Ia supernovae
\citep{ams+16}.



Unfortunately, the deep southern sky was not being optically surveyed
in unbiased fashion for transients and supernovae during the 2013--14
southern summer season. In particular, the Catalina Real-Time
Transient Survey (CRTS; \citealt{ddm+09}) had ceased use of its Siding
Springs Observatory 1\,m facility in July; the La Silla-QUEST
Supernova Survey (LSQ; \citealt{wbc+15}) was using a drift-scan
approach not suited (nor applied) to deep polar regions $\delta\simlt
-40\arcdeg$; and the Optical Gravitational Lensing Experiment
Real-Time Transient Search (OGLE-IV; \citealt{ogleiv}) was focused on
transient discovery and variable star studies within and near the
Small and Large Magellanic Clouds, without extending as far as the
location of \frb. Hence it is not possible to set quantitative limits
on the peak magnitude of any associated supernova to \frb\ beyond what
can be determined from the UVOT observations.

\subsection{X-Shooter Follow-Up}
\label{sub:xshooter}

We retrieved \xshooter\ \citep{xshooter} follow-up observations of the
two Swift \xray\ sources from the ESO Data Archive\footnote{ESO Data
  Archive: \newline
  \url{http://archive.eso.org/eso/eso_archive_main.html}}. Each
spectroscopic observation was preceded by four 180\,s $R$-band
acquisition images. We coadded the undithered exposures for each
target, yielding two 1\farcm 5 $\times$ 1\farcm 5 images with ${\rm
  FWHM} \approx 0.7\arcsec$ seeing, reaching to roughly $R\approx
22$\,mag as determined by photometry of unsaturated stars from the
\usnob\ catalog (Fig.~\ref{fig:sky}). The image of the region
surrounding \swiftxa\ recovers a likely point-source optical
counterpart seen in archival data, which has $R \approx
19.0$\,mag, coordinates R.A. \rasxgd{06}{43}{39}{84},
Dec.\ \dcsxgd{$-$51}{20}{46}{0}, and $r_{90}\approx 0\farcs 3$. The
image of the region surrounding \swiftxb\ reveals a single candidate
point-source optical counterpart with $R \approx 20.8$\,mag,
coordinates R.A. \rasxgd{06}{44}{09}{50},
Dec.\ \dcsxgd{$-$51}{18}{54}{0}, and $r_{90}\approx 0\farcs 3$
(Fig.~\ref{fig:sky}).


Each of the candidate optical counterparts to the two Swift
\xray\ sources was observed in two dithered 600\,s exposures through a
1\arcsec\ slit, yielding spectra from each of the three
\xshooter\ spectrographs (UVB, VIS, NIR). We retrieved the Phase~3
data products\footnote{ESO Phase 3 Archive Interfaces:
  \url{http://archive.eso.org/wdb/wdb/adp/phase3_main/form}}, which
provide the wavelength-calibrated, rectified, and coadded
two-dimensional spectral images and extracted (one-dimensional)
spectra for each spectrograph, as processed by the \xshooter\ Pipeline
\citep{xshpipe}. We examined the two-dimensional and extracted spectra
to determine the nature of each object.

The likely counterpart to \swiftxa\ (X1) exhibits broad, redshifted
emission lines of \ion{Mg}{2} (2798\AA) and H$\alpha$ with $z\approx
0.38$ and $\Delta v \approx 3500$\,\kmsec\ (FWHM), demonstrating its
nature as a quasar. A weaker, broad H$\beta$ emission line at this
redshift is also present. The redshift is confirmed and refined via
narrow emission lines of \fbion{O}{2} (3727\AA), \fbion{O}{3}
(4959\AA), \fbion{O}{3} (5007\AA), and \fbion{N}{2} (6584\AA), as
well as an absorption feature due to the \ion{Ca}{2} (3935\AA,
3970\AA) doublet. These yield a refined redshift of $z=0.383$,
\xray\ luminosity $L_{\rm X}\approx 7.2\times 10^{43}$\,\ergsec, and
absolute magnitude $M_R \approx -22.6$\,mag. This spectroscopic
identification confirms the source as the optical counterpart to
\swiftxa.

The likely counterpart to \swiftxb\ (X2) exhibits broad, redshifted
emission lines of \ion{C}{4} (1549\AA), \ion{Mg}{2} (2798\AA),
H$\beta$, and H$\alpha$ with $z\approx 1.53$ and $\Delta v \approx
3000$\,\kmsec\ (FWHM), demonstrating its nature as a quasar. The
redshift is confirmed via narrow emission lines of \fbion{O}{2}
(3727\AA), \fbion{O}{3} (4959\AA), and \fbion{O}{3} (5007\AA), which
yield a refined redshift of $z=1.525$, \xray\ luminosity $L_{\rm
  X}\approx 1.2\times 10^{45}$\,\ergsec, and absolute magnitude $M_R
\approx -24.5$\,mag. This spectroscopic identification confirms the
source as the optical counterpart to \swiftxb.
  

\subsection{Repeating Counterpart Constraints}
\label{sub:rptsearch}


Given our findings, and the known existence of repeating FRB sources,
we initiated a search for \gray\ activity in the directions of all
known FRBs regardless of relative timing. We established that none of
the 1050 triggered Swift GRBs have positions consistent with any of
the known FRBs. (We note here the previously published search for
Swift and Fermi counterparts to the repeating \frbrpt;
\citealt{ssh+16b}.)

We then examined the complete set of Swift BAT subthreshold events
from the Swift archive at HEASARC. We extracted all events that
occurred within $2\arcdeg$ of any FRB location, excluding FRB\,010621,
as it lies close to a bright \xray\ source. Data run from December
2004 to February 2016, excluding January 2011 (we identified
unresolved quality issues with data files for this month), giving a
total of 134 months. We searched for an excess of subthreshold events
within 15\arcmin\ of each FRB position by comparison to the larger
region around the FRB (and beyond 15\arcmin;
Table~\ref{tab:subth}). Estimated on-axis equivalent exposure times
for each FRB position, derived from the exposure map in
\citet{batsurvey70}, are listed. We find no statistically-significant
excess for any of the examined positions, and set a 3$\sigma$ upper
limit on the number of such subthreshold events of $<$14 for the
typical such position (corresponding to rates per Ms of on-axis
equivalent exposure of $<$0.73\,Ms$^{-1}$), and $<$11 for
\frbrpt\ ($<$0.75\,Ms$^{-1}$). A stacked search over all 15
non-repeater FRB positions gives a limit of $<$29 excess subthreshold
events in all or $<$2 per FRB position ($<$0.11\,Ms$^{-1}$).


\begin{table}
\footnotesize

  \caption{\label{tab:subth}%
    Swift BAT subthreshold events near FRB positions}

  \begin{center}

  \begin{tabular}{l r c c r r c}

   \hline \hline

   \mccc{FRB} & 
     Exp. & 
     \mcc{2}{$n_{\rm obs}$} & 
     $n_{\rm exp}$ & 
     $n_{\rm max}$ & 
     $r_{\rm max}$  \\ \cline{3-4}

   ~ & (Ms) 
     & ($<$2\arcdeg)
     & ($<$15\arcmin) & ~ & ~
     & (Ms$^{-1}$) \\ \hline

   121102 & 14.5 & 326 & \phn 4 &  5.1 & 10.9 & 0.75 \\ 
   010125 & 15.3 & 672 &     10 & 10.5 & 14.5 & 0.95 \\
   010724 & 23.9 & 657 &     12 & 10.2 & 17.6 & 0.74 \\
   090625 & 23.9 & 756 &     12 & 11.8 & 16.0 & 0.67 \\
   110220 & 14.5 & 450 & \phn 7 &  7.0 & 13.6 & 0.94 \\
   110523 & 15.3 & 470 &     11 &  7.3 & 19.1 & 1.25 \\
   110626 & 14.5 & 540 & \phn 4 &  8.5 &  7.5 & 0.52 \\
   110703 & 15.3 & 501 & \phn 9 &  7.8 & 15.8 & 1.03 \\
   120127 & 15.3 & 519 & \phn 6 &  8.1 & 11.0 & 0.72 \\
   121002 & 21.1 & 827 &     10 & 13.0 & 12.0 & 0.57 \\
   130626 & 12.4 & 515 & \phn 9 &  8.0 & 15.5 & 1.25 \\
   130628 & 16.3 & 529 & \phn 8 &  8.3 & 13.8 & 0.85 \\
   130729 & 14.5 & 503 & \phn 3 &  7.9 &  6.4 & 0.44 \\
   131104 & 24.9 & 818 &     14 & 12.8 & 17.8 & 0.72 \\
   140514 & 14.5 & 451 & \phn 7 &  7.0 & 13.6 & 0.94 \\
   150418 & 17.2 & 589 & \phn 6 &  9.3 &  9.9 & 0.57 \\  \hline

   \multicolumn{7}{p{3.2in}}{\scriptsize {\bfseries Note---}Equivalent
     on-axis exposure times in Ms (Exp.) are estimated from Fig.~1 in
     \citet{batsurvey70}\ and renormalized to the 134 months of our
     search. $n_{\rm obs}$ is the number of subthreshold events within
     the specified distance (2$\arcdeg$ or 15$\arcmin$) from each FRB,
     $n_{\rm exp}$ is the expected number of subthreshold events
     expected within 15$\arcmin$ based on the number within 2$\arcdeg$
     (excluding those within 15$\arcmin$), $n_{\rm max}$ is the
     3$\sigma$ upper limit on the number of excess FRB-associated
     subthreshold events that can be accommodated given the
     observations, and $r_{\rm max}$ is the 3$\sigma$ upper limit on
     the rate of such FRB-associated subthreshold events per Ms
     on-axis equivalent exposure. FRB\,010621 is excluded from this
     analysis, as it lies close to a bright X-ray source. \frbrpt\ is
     a known repeating FRB source \citep{ssh+16a}.}

   \end{tabular}
  \end{center}

\end{table}


\section{Discussion}
\label{sec:discuss}


The observation of an energetic \gray\ counterpart to \frb\ challenges
FRB models. Nearby Galactic ($D\simlt 1$\,kpc) flare stars have been
proposed as repeating FRB sources \citep{lsm14}. \frb, however, has no
apparent variable stars within its sky localization \citep{maoz+15},
and has not been observed to repeat \citep{rsj15}. A local
extragalactic origin, as from giant pulses of a pulsar in the Carina
dwarf spheroidal galaxy or Magellanic stream \citep{rsj15}, is also
excluded, as pulsar giant pulses are not accompanied by
\gray\ emission.

Extragalactic FRBs from hyperflares of young magnetars \citep{pp10},
or from rapidly-rotating pulsars or magnetized white dwarfs
\citep{cw16,mkm16}, might be seen repeatedly from sources at $z\simlt
0.1$, with excess local dispersion produced by a surrounding dense
environment such as a wind nebula, supernova ejecta, or the central
high-density regions of star-forming galaxies
\citep{mls+15,mkm16,piro16}. In magnetar scenarios this might bring
\frb\ close enough ($D\simlt 160$\,Mpc) to accommodate the maximum
expected magnetar flare energy of $E_\gamma\sim 10^{49}$\,erg, which
can be achieved if inner magnetic field strengths are $B_c\sim
10^{16}$\,G. However, its \gray\ emission timescale is
$\simgt$500$\times$ longer than observed for the \sgrbig\ hyperflare
($T_{90} = 0.2$\,s; \citealt{hbs+05}).

For all such post-SN or post-merger progenitors, extreme energies
$E_\gamma\simgt 10^{49}$\,erg (whether powered by internal magnetic
fields or spindown), unless accompanied by beaming, suggest young ages
$\tau\simlt 100$\,yr. At the same time, the absence of significant
free-free absorption along the line of sight requires the surrounding
ejecta to be relatively dispersed, $\tau\simgt 10$ to 100\,yr
\citep{mkm16}. It is not clear whether these requirements can be met
simultaneously.


Relatively long-duration ($T_{90}\simgt 30$\,s), smoothly-evolving
\gray\ transients are generic to the relativistic shock breakouts that
accompany a variety of massive stellar deaths \citep{kbw10,ns12}, and
so might occur at a rate comparable to the 6700\,\gpcyr\ FRB rate.
The thick stellar wind surrounding collapsing massive stars would act
as a local source of dispersion (reducing distances, while increasing
the inferred volumetric rate); however, the density at the point of
shock breakout is large enough that it is not clear how the FRB would
not be immediately quenched by free-free and other absorption
processes \citep{kon+14,mkm16}.
For an energy output $E_\gamma \approx 4\times 10^{49}$\,erg, as seen
from the two Swift shock breakout events GRBs\,060218 and 100316D
\citep{ns12}, \frb\ would be at $D = 290^{+90}_{-50}$\,Mpc ($z \approx
0.07\pm 0.02$). This distance is (roughly) sufficient to keep the
\xray\ afterglows of these two GRB-SNe below the flux limit of the XRT
observations for \frb.


Relativistic tidal disruption events from otherwise quiescent galaxies
form another population of long-duration \gray\ transients seen by
Swift \citep{bkg+11,bgm+11,zbs+11,ckh+12,bls+15}; however, these are
rare, with an observed rate of $10^{-3}
f_{b\rm{,TDE}}^{-1}$\,\gpcyr\ \citep{bls+15}, where the beaming factor
is uncertain but typically taken to be $f_{b,{\rm TDE}} \ge 10^{-4}$.
Moreover, the coherent radio emission mechanism that would generate
FRBs in this case is not clear.


FRBs from binary neutron star (BNS) coalescence \citep{totani13}, the
subsequent collapse of an overmassive neutron star to a black hole
\citep{fr14,zhang14}, or black hole plus neutron star (\mbox{BH-NS})
coalescence \citep{mll15} might yield coincident high-energy emission
as a short-hard gamma-ray burst ($T_{90}\simlt 2$\,s;
\citealt{ffp+05}), or subsequently as extended \citep{kbg+15} or
flaring \citep{zhang14} emission during the post-burst phase. If the
prompt \gray\ emission were missed due to beaming, yet accompanied by
longer-lasting and less-luminous off-axis emission as in ``orphan
afterglow'' scenarios (e.g., \citealt{gsc+15}), this would be
consistent with the absence of short burst-like emission for \frb;
coherent radio emission could be produced immediately pre-merger,
although details of the emission mechanism are uncertain. Using the
\gray\ duration constraint and lightcurve data for \swiftgrt, we
constrain the start time of \gray\ emission to $\Delta t \le 215$\,s
after the FRB in this case.

The FRB volumetric rate lower limit of 6700\,\gpcyr\ is consistent
with the upper range of allowed BNS + \mbox{BH-NS} merger rates
\citep{ckw16}, and with the upper end of beaming-corrected estimates
of the short-hard GRB rate \citep{chp+12}. These merger rates should
be quickly refined or reduced via upcoming runs of the advanced
gravitational wave facilities \citep{ckw16}.

Assuming a beaming-corrected short-hard GRB rate of $R_{\rm SHB}\sim
1000$\,\gpcyr, then, based on the sensitivity of BAT with the imaging
trigger, the beaming factor of any associated extended emission must
be $f_{b{\rm, SHB-EE}} \simlt 3\times{10}^{-2}$, which is not much
larger than the estimated beaming factor of the prompt emission,
$f_{b{\rm, SHB}}\sim {10}^{-2}$ \citep{chp+12}. Thus, the BNS scenario
for FRBs and short-hard GRBs predicts that coincidences between FRBs
and short gamma-ray bursts should be almost as common as the
``off-axis''-type coincidence that could be invoked to explain \frb.


The large gamma-ray energy of $E_\gamma \sim 5\times 10^{51}$\,erg is
encouraging for detection of counterparts with late-time follow-up
observations.  If accompanied by a relativistic jet with comparable
kinetic energy, it implies that FRB \xray\ and radio afterglows can be
detected as with short-hard GRBs.

If some FRBs are accompanied by \gray\ transients with supernova or
luminous afterglow counterparts, then these offer the prospect of
arcsecond localizations, host galaxy identifications, and distance
measurements for FRBs, and hence may facilitate application of FRBs to
outstanding questions of fundamental physics (including tests of
Lorentz invariance and the equivalence principle) and cosmology
\citep{ioka03,inoue04,mcquinn14,wgw+15,wzg+16,arg16}.


Had it been viewed at close to normal incidence ($>$90\% coding),
\swiftgrt\ would have had $\mathcal{S}\simgt 20\sigma$, almost
certainly yielding a BAT image trigger. Past Swift image trigger
events include the relativistic shock breakout GRB-SNe and
relativistic tidal disruption events already mentioned, along with
multiple additional long \citep{swl+11} and ``ultra-long'' bursts
\citep{lts+14}, which have been argued to represent a distinct class
of GRBs.


This presents a puzzle of interpretation because of the profound
mismatch in inferred rates between the known varieties of
long-duration \gray\ transient -- we estimate $\sim$25 yr$^{-1}$ above
BAT threshold -- and the FRB all-sky rate of 2100 day$^{-1}$
\citep{cpk+16}. While we cannot hope to resolve this mismatch with
just one counterpart and one upper limit among two non-repeating FRBs
observed (Table~\ref{tab:batprompt}), we do note the following: (1)
The existence of the repeating \frbrpt, combined with limits on
repetition from other FRBs including \frb, already suggests that
multiple source populations contribute to the FRB phenomenon; (2) The
limits on repeating \gray\ transients derived here strongly suggest
that repeating-type FRBs do not have \gray\ counterparts; (3) The long
timescale of \swiftgrt\ acts efficiently to hide it from satellite
observatories other than Swift; and (4) Gamma-ray-bright FRBs may be a
minority even among those that generate \gray\ emissions.


\section{Conclusions}
\label{sec:conclude}


Our discovery that some FRBs are accompanied by energetic
\gray\ transients dramatically alters the basic picture of these
events. To date the FRBs have excited interest on the basis of their
short timescales, likely-cosmological distances, and high event rates;
however, received fluences of the events have been small, implying
potentially modest energy requirements ($E_{\rm radio} \sim 4\times
10^{41}$\,erg in this case). The \gray\ energy requirement for
\frb\ is more than $10^{9}$ times greater, with dramatic implications
for source models and a substantial improvement in the prospects for
long-lived counterparts, including \xray\ and radio afterglows. The
increased energy scale also raises the consequences of the FRB
phenomenon for the bursts' host galaxies and their evolution, as well
as (for relatively young progenitors) the evolution of their
star-forming regions.

Looking forward, this result should further energize efforts aimed at
real-time discovery and multiwavelength follow-up observations of
FRBs. We expect these searches will now routinely extend into the
subthreshold regime for wide-field experiments such as Swift, Fermi,
and the High Altitude Water Cherenkov facility (HAWC;
\citealt{hawcdesign}); and to multimessenger facilities including
IceCube \citep{icfacility}, ANTARES \citep{antares}, Pierre Auger
\citep{auger}, and the gravitational wave observatories
\citep{lvc12}. Such a joint and coordinated search for counterparts
would be a natural project for the Astrophysical Multimessenger
Observatory Network (AMON; \citealt{amon}) now under construction at
Penn State. The most promising immediate prospects probably relate to
rapid and sustained \xray\ and radio afterglow and nearby ($z\simlt
0.1$) supernova searches in the wake of each bright and well-localized
FRB.


While our finding resolves a pressing question regarding the FRBs --
do they exhibit non-radio counterparts -- and yields important clues
as to the nature of this event, it simultaneously bolsters the case
for multiple FRB source populations. These likely wait to be revealed
by future radio searches and rapid-response follow-up observations
across the electromagnetic spectrum and via multimessenger facilities.



\acknowledgments

We gratefully acknowledge support from Penn State's Office of the
Senior Vice President for Research, the Eberly College of Science, and
the Penn State Institute for Gravitation and the Cosmos.
K.M. acknowledges support by NSF Grant No. PHY-1620777, while
P.M. acknowledges NASA support via grant NNX13AH50G. This research has
made use of data and software provided by the High Energy Astrophysics
Science Archive Research Center (HEASARC), which is a service of the
Astrophysics Science Division at NASA/GSFC and the High Energy
Astrophysics Division of the Smithsonian Astrophysical Observatory. It
has also made use of data and online tools supplied by the UK Swift
Science Data Centre at the University of Leicester, and of data
obtained from the ESO Science Archive Facility.

\facilities{Swift (BAT, XRT, UVOT), VLT (X-Shooter)}

\software{HEASOFT, IDL, SciPy, CIAO}



\bibliographystyle{apj_8}


\end{document}